\documentclass[apj]{emulateapj}

\usepackage{natbib}
\shorttitle{ALMA molecular gas observations of HD\,21997}
\shortauthors{K\'osp\'al et al.}

\begin{document}

\title{ALMA observations of the molecular gas in the debris disk of\\
  the 30 Myr old star HD\,21997}

\author{\'A. K\'osp\'al\altaffilmark{1,11}}
\author{A. Mo\'or\altaffilmark{2}}
\author{A. Juh\'asz\altaffilmark{3}}
\author{P. \'Abrah\'am\altaffilmark{2}}
\author{D. Apai\altaffilmark{4}}
\author{T. Csengeri\altaffilmark{5}}
\author{C. A. Grady\altaffilmark{6,7}}
\author{Th. Henning\altaffilmark{8}}
\author{A. M. Hughes\altaffilmark{9}}
\author{Cs. Kiss\altaffilmark{2}}
\author{I. Pascucci\altaffilmark{10}}
\author{M. Schmalzl\altaffilmark{3}}

\altaffiltext{1}{European Space Agency (ESA-ESTEC, SRE-SA), P.O.
  Box 299, 2200AG, Noordwijk, The Netherlands; akospal@rssd.esa.int}

\altaffiltext{2}{Konkoly Observatory, Research Centre for Astronomy
  and Earth Sciences, Hungarian Academy of Sciences, P.O. Box 67, 1525
  Budapest, Hungary}

\altaffiltext{4}{Department of Astronomy and Department of Planetary
  Sciences, The University of Arizona, Tucson, AZ 85721, USA}

\altaffiltext{5}{Max-Planck-Institut f\"ur Radioastronomie, Auf dem
  H\"ugel 69, D-53121 Bonn, Germany}

\altaffiltext{6}{NASA Goddard Space Flight Center, Code 667,
  Greenbelt, MD 20771, USA}

\altaffiltext{7}{Eureka Scientific, 2452 Delmer Street, Suite 100,
  Oakland, CA 94602, USA}

\altaffiltext{8}{Max-Planck-Institut f\"ur Astronomie, K\"onigstuhl 17,
  D-69117 Heidelberg, Germany}

\altaffiltext{9}{Astronomy Department, Wesleyan University,
  Middletown, CT 06459, USA}

\altaffiltext{3}{Leiden Observatory, Leiden University, Niels Bohrweg
  2, NL-2333 CA Leiden, The Netherlands}

\altaffiltext{10}{Lunar and Planetary Laboratory, Department of
  Planetary Sciences, University of Arizona, 1629 East University
  Boulevard, Tucson, AZ 85721, USA}

\altaffiltext{11}{ESA Fellow.}


\begin{abstract}
The 30 Myr old A3-type star HD\,21997 is one of the two known debris
dust disks having a measurable amount of cold molecular gas. With the
goal of understanding the physical state, origin, and evolution of the
gas in young debris disks, we obtained CO line observations with the
Atacama Large Millimeter/submillimeter Array (ALMA). Here we report on
the detection of $^{12}$CO and $^{13}$CO in the $J$=2--1 and $J$=3--2
transitions and C$^{18}$O in the $J$=2--1 line. The gas exhibits a
Keplerian velocity curve, one of the few direct measurements of
Keplerian rotation in young debris disks. The measured CO brightness
distribution could be reproduced by a simple star+disk system, whose
parameters are $r_{\rm in} < 26$\,AU, $r_{\rm
  out}\,{=}\,138\,{\pm}\,20$\,AU, $M_*=1.8^{+0.5}_{-0.2}\,M_{\odot}$,
and $i\,{=}\,32\fdg6\,{\pm}\,3\fdg1$. The total CO mass, as calculated
from the optically thin C$^{18}$O line, is about
(4--8)$\times$10$^{-2}\,M_{\oplus}$, while the CO line ratios suggest
a radiation temperature on the order of 6--9\,K. Comparing our results
with those obtained for the dust component of the HD\,21997 disk from
the ALMA continuum observations by Mo\'or et al., we conclude that
comparable amounts of CO gas and dust are present in the
disk. Interestingly, the gas and dust in the HD\,21997 system are not
co-located, indicating a dust-free inner gas disk within 55\,AU of the
star. We explore two possible scenarios for the origin of the gas. A
secondary origin, which involves gas production from colliding or
active planetesimals, would require unreasonably high gas production
rates and would not explain why the gas and dust are not
co-located. We propose that HD\,21997 is a hybrid system where
secondary debris dust and primordial gas coexist. HD\,21997, whose age
exceeds both the model predictions for disk clearing and the ages of
the oldest T\,Tauri-like or transitional gas disks in the literature,
may be a key object linking the primordial and the debris phases of
disk evolution.
\end{abstract}



\section{Introduction}
\label{sec:intro}

Nearly all young stars harbor circumstellar disks whose thermal
emission produces a strong infrared excess with fractional
luminosities of $L_{\rm d} / L \geq 0.1$. At the early stage of their
evolution, the masses of these disks (typically a few percent of the
mass of the central star) are dominated by gas with only a few percent
of the mass in small, submicron-sized dust grains. Both the dust and
the gas are of primordial origin. As the disk evolves, dust settles in
the midplane and the grains eventually form larger bodies,
planetesimals and planets. The gas is removed by viscous accretion
\citep[e.g.,][]{lyndenbell1974}, by photoevaporation
\citep[e.g.,][]{alexander2008}, or by planet formation. Current
observational results imply that the primordial gas is mostly depleted
at ages of $\lesssim$10 Myr \citep{pascucci2006, fedele2010}.

A detectable amount of infrared excess is also present in many older,
main-sequence stars, with typical fractional luminosities of $L_{\rm
  d} / L \leq 10^{-3}$ \citep[e.g.,][]{roccatagliata2009}. These
debris disks are fundamentally different from primordial disks. Their
masses, as inferred from the emission of small dust grains, are
usually below 1\,$M_{\earth}$ and they are practically
gas-free. Without the stabilizing effect of surrounding gas, the
lifetime of individual dust grains in debris disks is very short due
to removal by dynamical interactions with stellar radiation. Thus,
dust needs to be continuously replenished by collisions and/or
evaporation of previously formed planetesimals
\citep[e.g.,][]{wyatt2008}. The same processes would in principle
produce gas as well, via sublimation of planetesimals
\citep{lagrange1998}, photodesorption from dust grains
\citep{grigorieva2007}, vaporization of colliding dust particles
\citep{czechowski2007}, or collision of comets or icy planetesimals
\citep{zuckerman2012}. Thus, the secondary gas that may be present in
a debris disk is dominated by CO and H$_2$O \citep{mumma2011} and only
a small amount of H$_2$, mainly originating from the dissociation of
H$_2$O, is expected.

In theory, it is possible that systems in transition between the
primordial and the debris state possess hybrid disks, i.e.~primordial
gas is accompanied by secondary dust. It is also possible that the
outer disk is still primordial, while the inner disk is already
composed of secondary material \citep{wyatt2008, krivov2009}. So far,
only six debris disks with detectable gas component are known. The
edge-on orientation of the disks around $\beta$\,Pic and HD\,32297
allow the detection of a very small amount of circumstellar gas based
on the presence of absorption lines \citep{roberge2000,
  redfield2007}. Another member of the $\beta$\,Pic moving group,
HD\,172555, also contains some gas as evidenced by [O\,I] emission at
63$\,\mu$m \citep{riviere2012}. Recently, the [CII] line at
158$\,\mu$m was detected in $\sim$30 Myr old HD\,32297
\citep{donaldson2013}. A substantial amount of molecular gas has been
detected at millimeter wavelengths in the debris disk around the young
main-sequence stars 49\,Ceti \citep{hughes2008} and HD\,21997
\citep{moor2011b}. The age of these systems are between 10 and
40\,Myr, which partly overlaps with the transition period from the
primordial to the debris phases. Although HD\,21997 and 49\,Cet are
the oldest members of the gaseous debris disk sample, they also
contain the largest amount of cold gas. Thus, it is worth considering
whether in their case the detected gas is of primordial or secondary
origin.

In this paper, we focus on HD\,21997, which is an A3-type star at a
distance of 72\,pc (based on Hipparcos parallax;
\citealt{vanleeuwen2007}), belonging to the well-dated 30\,Myr old
Columba moving group \citep{moor2006,torres2008}. In an earlier study,
we detected molecular gas in the $J$=3--2 and $J$=2--1 transitions
(CO(3--2) and CO(2--1)) with the APEX telescope
\citep{moor2011b}. Motivated to understand the origin and evolutionary
status of the gas component in this system, we obtained
(sub)millimeter interferometric continuum and CO line observations
with the Atacama Large Millimeter/submillimeter Array (ALMA). Our aim
was to spatially resolve the disk in order to study the relative
location of the gas and dust components, determine the distribution of
the different CO isotopologues, precisely measure the gas mass, and
analyze gas kinematics. In this paper, we present our results on the
gas content of the disk, while the dust continuum observations are
analyzed in \citet[][hereafter, Paper\,I]{moor2013}.


\section{Observations and data reduction}

We observed HD\,21997 with ALMA in Cycle 0 using a compact
configuration (PI: \'A.~K\'osp\'al). The quasar J0403$-$360 served as
a bandpass and gain calibrator, while Callisto was used to set the
absolute amplitude scale using the CASA Butler-JPL-Horizons 2010
model. We obtained both line and continuum data. Here, we focus on the
line observations, while the continuum measurements are presented in
Paper\,I.

In Band 6, we targeted the (2--1) transitions of $^{12}$CO, $^{13}$CO,
and C$^{18}$O. We used the frequency division correlator mode
(FDM). Each of the four simultaneously observed spectral windows
offered 3840 channels with a channel separation of 122\,kHz, resulting
in a bandwidth of 469\,MHz per window. The spectral resolution
corresponds to a velocity resolution of 0.33\,km\,s$^{-1}$ at
230\,GHz. Due to restrictions in the setup of the local oscillators,
the three CO lines could not be observed simultaneously; therefore,
two different correlator configurations were prepared. One setup (A1)
was configured to observe $^{13}$CO(2--1) and C$^{18}$O(2--1), and it
was executed on 2011 November 29 using 14 antennas, with baselines
ranging from 9.7 to 148\,k$\lambda$. The other setup (A2) was tuned to
have $^{12}$CO(2--1) in one of the spectral windows, and it was
observed on 2011 December 31, using 16 antennas, with baselines
between 9.6 and 196\,k$\lambda$. Each of these two setups had an
on-source time of 49.4\,minutes.

In Band 7, we targeted the (3--2) transitions of $^{12}$CO and
$^{13}$CO, as well as the (7--6) transition of CS. The correlator was
set up in FDM with 3840 channels, a channel separation of 244\,kHz,
and a total bandwidth of 938\,MHz. The spectral resolution corresponds
to a velocity resolution of 0.44\,km\,s$^{-1}$ at 345\,GHz. Again, two
correlator configurations were used. In one setup (B1),
$^{12}$CO(3--2) and $^{13}$CO(3--2) were targeted. Observations were
done on 2011 November 3 and 4 using 14 and 15 antennas, respectively,
with baselines ranging from 14 to 154\,k$\lambda$. The other setup
(B2) covered the CS(7--6) line, and was observed on 2011 November 3
with 14 antennas and baselines between 14 and 153\,k$\lambda$. The
rest frequency of CS(7--6) also allowed us to obtain additional
coverage of $^{13}$CO(3--2) in this configuration. Both configurations
had an on-source time of 49.4\,minutes, resulting in
$\sim$100\,minutes of observing time for $^{13}$CO(3--2).

We subtracted the continuum emission in the spectral cubes using the
CASA task {\tt uvcontsub}, and individually cleaned the spectral
region around each molecular line using the CASA task {\tt clean} with
robust weighting. A summary of the observing setup, beam size and
position angle (P.A.) for each line, and the rms noise measured in the
channel maps is presented in Table~\ref{tab:results}.


\section{Results}

\subsection{Integrated Line Profiles}

\begin{figure}[h!]
\epsscale{1.18}
\plotone{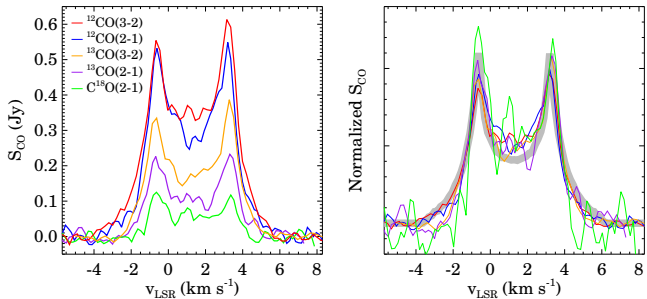}
\caption{CO spectra of HD\,21997 observed with ALMA. The left panel
  shows integrated flux densities in Jy, while the right panel
  displays spectra normalized to their respective line areas, to
  better show the similarities in the line profiles. In the right
  panel, an additional thick gray line shows the best-fit model
  described in Section~\ref{sec:model}.
\label{fig:cospectra}}
\end{figure}

We detected all targeted CO lines, i.e., $^{12}$CO(2--1),
$^{12}$CO(3--2), $^{13}$CO(2--1), $^{13}$CO(3--2), and
C$^{18}$O(2--1). Neither the CS(7--6) line nor any other lines are
detected in our ALMA data cubes. In order to obtain integrated
profiles for the CO lines, we added all the flux within a radius of
2$\farcs$8 of the stellar position (the smallest aperture that
contains the whole flux of the object) for each velocity channel. The
resulting line profiles are plotted in Figure~\ref{fig:cospectra}
(left panel). In line with our earlier APEX observations
\citep{moor2011b}, the lines are all double-peaked, typical for
emission arising from rotating material. The intensities of the two
$^{12}$CO lines are also consistent within the uncertainties with our
earlier single-dish APEX results, indicating that there is no
significant flux loss due to the interferometer filtering out
large-scale structures. The peak and central velocities are the same
for all lines. Within the uncertainties, the profiles are symmetric
around a systemic velocity of $v_{\rm sys}$=1.29\,km\,s$^{-1}$
(measured in the local standard of rest (LSR) system), which agreeswith the radial velocity of the central star. The intensity-weighted
average velocity for all lines is the same
(1.29\,$\pm$\,0.05\,km\,s$^{-1}$) as well. The symmetric profiles
suggest that the rotating material is distributed axisymmetrically
around the central star.

Table~\ref{tab:results} presents the line fluxes integrated from
$-4$\,km\,s$^{-1}$ to 6.6\,km\,s$^{-1}$, a velocity interval that
covers the whole line, as well as the peak fluxes. The strongest CO
line is the 3--2 transition of $^{12}$CO, followed by the 2--1 line of
the same isotopologue, then the 3--2 and 2--1 transitions of
$^{13}$CO, while the 2--1 line of C$^{18}$O is the faintest among our
detections. If we normalize the line profiles
(Figure~\ref{fig:cospectra}, right panel), we find that the lines have
very similar profiles. We checked which velocity channels contain
significant emission above the 3$\sigma$ level, and found that these
span a range of $\pm$4.3\,km\,s$^{-1}$ around the systemic velocity.

\begin{table*}
\begin{center}
{\scriptsize
\caption{Rest Frequencies, Observing Setups (See Text For Details),
  Beam Sizes and Position Angles, Noise Levels, and Line Fluxes for
  the CO Observations of HD\,21997\label{tab:results}}
\begin{tabular}{ccccc@{}cccc}
\tableline\tableline
Line           & Frequency & Setup  & Beam Size        & Beam PA      & Noise             & Peak Flux & Total Flux        & Total Flux \\
               & (GHz)     &        & ($''\times''$)   & ($^{\circ}$) & (mJy\,beam$^{-1}$\,channel$^{-1}$) & (Jy)      & (Jy\,km\,s$^{-1}$) & (W\,m$^{-2}$) \\
\tableline
$^{12}$CO(3--2) & 345.796   & B1     & 1.19$\times$1.47 & 117.5        & 7.9 & 0.615     & 2.52$\,\pm\,$0.27 & (2.91$\,\pm\,$0.31)$\times$10$^{-20}$  \\
$^{12}$CO(2--1) & 230.538   & A2     & 1.14$\times$1.21 & 16.4         & 8.6 & 0.566     & 2.17$\,\pm\,$0.23 & (1.67$\,\pm\,$0.18)$\times$10$^{-20}$  \\
$^{13}$CO(3--2) & 330.588   & B1, B2 & 1.20$\times$1.38 & $-$49.9      & 5.7 & 0.387     & 1.35$\,\pm\,$0.15 & (1.49$\,\pm\,$0.17)$\times$10$^{-20}$  \\
$^{13}$CO(2--1) & 220.399   & A1     & 1.19$\times$1.92 & 19.0         & 7.6 & 0.238     & 0.82$\,\pm\,$0.10 & (6.04$\,\pm\,$0.74)$\times$10$^{-21}$  \\
C$^{18}$O(2--1) & 219.560   & A1     & 1.21$\times$1.92 & 18.6         & 9.0 & 0.129     & 0.36$\,\pm\,$0.06 & (2.63$\,\pm\,$0.44)$\times$10$^{-21}$  \\
\tableline
\end{tabular}
}
\end{center}
\end{table*}

\subsection{Spatially Resolved CO Emission}
\label{sec:resolved}

\begin{figure}[h!]
\epsscale{0.61}
\plotone{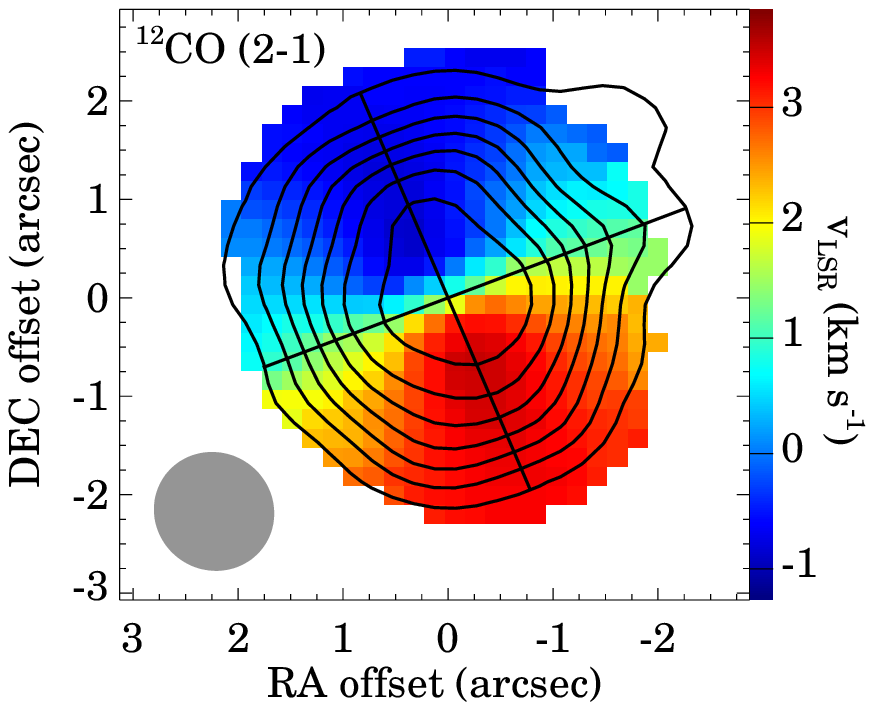}
\epsscale{0.51}
\plotone{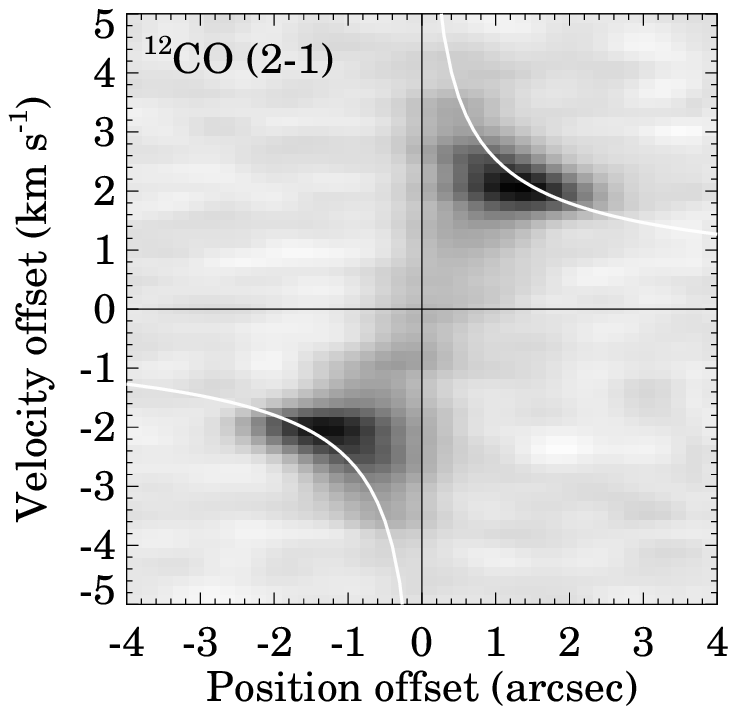}
\caption{Left panel: zeroth moment (with contours) and first moment
  (with color scale) of the $^{12}$CO (2--1) emission. The major and
  minor axes are plotted with straight black lines. The gray ellipse
  in the lower left corner indicates the beam. Right panel:
  position-velocity diagram of $^{12}$CO (2--1) emission along the
  major axis. Overplotted in white is a Keplerian rotation curve for
  $M_*=1.8\,M_{\odot}$ and $i=32\fdg$6 (see Section~\ref{sec:model}).
\label{fig:pvd}}
\end{figure}

The disk was spatially resolved in all CO transitions. We calculated
zeroth and first moment maps for each data cube (for an example, see
Figure~\ref{fig:pvd}, left panel). Figure~\ref{fig:pvd} shows that the
intensity profile peaks at the stellar position and smoothly decreases
with radial distance. The velocity maps indicate that to the north of
the star, material is approaching, while in the south, material is
receding, with respect to $v_{\rm sys}$. The same characteristics can
be seen for the other lines as well. Assuming rotating disk
kinematics, we can determine the rotation axis by fitting a line to
those pixels in the first moment map where the velocity equals to
$v_{\rm sys}$. This axis also marks the minor axis of the inclined
disk image. We determined the P.A.~of the major axis on each image and
obtained on average 22$\fdg$6$\pm$0$\fdg$5. The extent of the emission
along the minor and major axes defined by the 3$\sigma$ contour in
Figure~\ref{fig:pvd} (left panel) is 3$\farcs$5$\times$4$\farcs$0.

\begin{figure*}[h!]
\hspace*{-18mm}
\epsscale{0.738}
\plotone{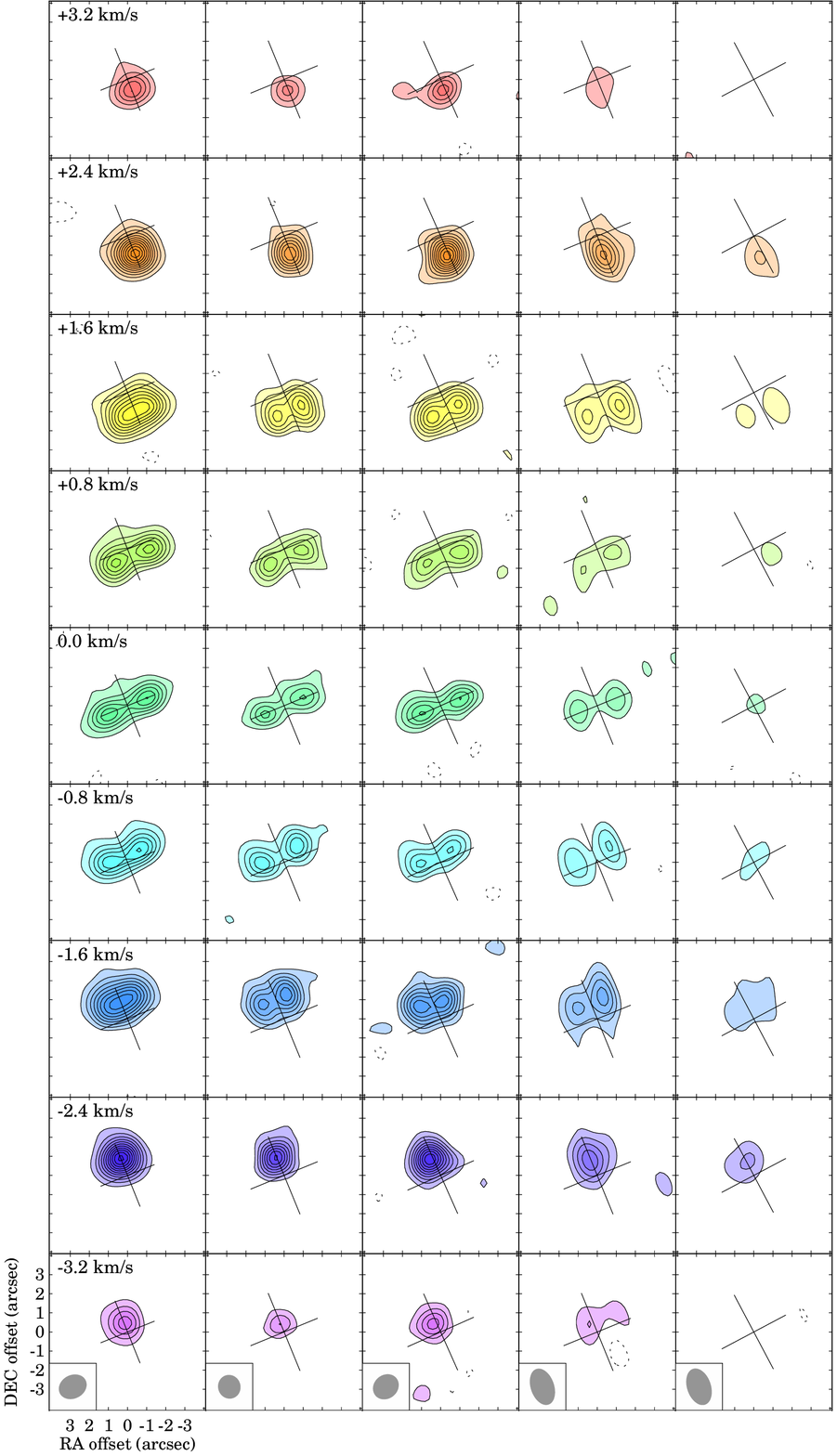}
\hspace*{-20mm}
\epsscale{0.385}
\plotone{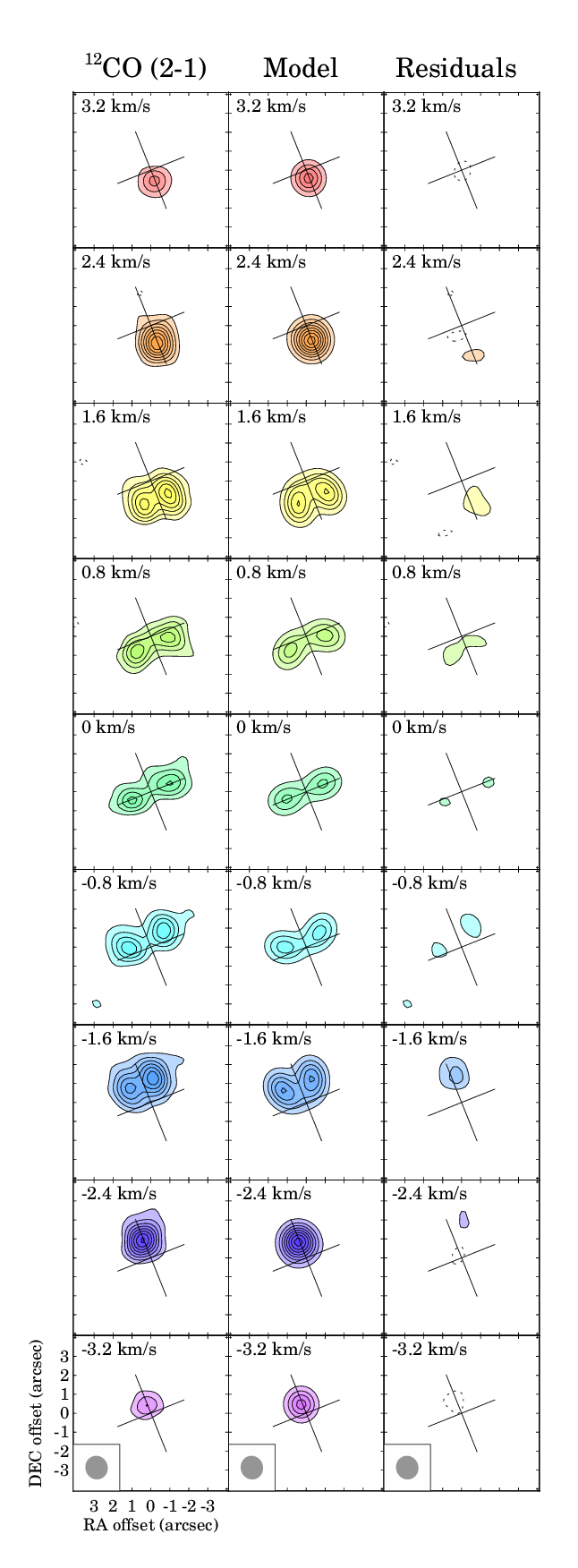}
\caption{Left: channel maps of CO observations of
  HD\,21997. Velocities indicated in the upper left corners are
  relative to the systemic velocity. Solid contours mark the
  3$\sigma$, 6$\sigma$, 9$\sigma$, 12$\sigma$, etc... levels, while
  dashed contours indicate the $-$3$\sigma$ level. Right: observed and
  modeled CO channel maps for the $^{12}$CO (2--1) line, and
  residuals. Details of the model are described in
  Section~\ref{sec:model}. Contours are the same as in the left panel.
\label{fig:channelmaps}}
\end{figure*}

Figure~\ref{fig:pvd} (right panel) shows a position velocity diagram
measured along the major axis of the disk marked in
Figure~\ref{fig:pvd} (left panel). Most of the emission comes from two
well-separated regions. The highest velocities are observed closest to
the stellar position, and velocity gradually decreases with increasing
distance, a pattern characteristic of Keplerian-like disk rotation. In
Figure~\ref{fig:channelmaps} (left panel) we plot the observed CO
emission in nine different velocity channels between $-$3.2 and
3.2\,km\,s$^{-1}$ relative to $v_{\rm sys}$, in steps of
0.8\,km\,s$^{-1}$. The plots show that the emission in the most
extreme velocity channels is rather compact and is located close to
the stellar position. Toward lower velocities, the peak of the
emission appears farther from the stellar position. For velocities
close to the systemic velocity, the emission is double peaked. The
morphology of the images is very similar on either side of the
systemic velocity, apart from a mirroring. It is evident from the
plots that the different CO lines exhibit the same emission pattern,
allowing for the differences in the beam shape/size and
the signal-to-noise ratio.

\subsection{Temperature and Mass of the CO Gas}

Despite their very different abundances, the fact that the intensities
of the three different CO isotopologues are of the same order of
magnitude already indicates that $^{12}$CO is probably optically
thick. We observed the (2--1) transition for three different CO
isotopologues. If we denote the optical depths of $^{12}$CO,
$^{13}$CO, and C$^{18}$O by $\tau_{12}$, $\tau_{13}$, and $\tau_{18}$,
respectively, then the ratio of the $^{12}$CO to the C$^{18}$O line
can be approximated as $(1-e^{-\tau_{12}})/(1-e^{-\tau_{18}})$ and a
similar formula holds for the $^{13}$CO to the $^{18}$CO line
ratio. Assuming that the optical depths of the different isotopologues
follow the same proportions as the abundance ratios typical of local
interstellar matter \citep{wilson1994}, then $\tau_{12} = 560
\tau_{18}$, and $\tau_{13} = 7.4 \tau_{18}$. Using these numbers, we
obtained $\tau_{18}=0.2$ from the $^{12}$CO to the C$^{18}$O line
ratio, and $\tau_{18}=0.6$ from the $^{13}$CO to the C$^{18}$O line
ratio. This result means that $\tau_{13}$ is in the 1.5--4.5 range,
while $\tau_{12}$ is between about 100 and 300. Using comet-like
isotopic ratios would yield similar results
\citep{eberhardt1995,bockelee2012,rousselot2012}. According to
\citet{visser2009}, and references therein, the $^{12}$C/$^{13}$C
ratio may be a factor of two smaller or larger, while the
$^{16}$O/$^{18}$O ratio may be three to five times larger than the
local elemental ratios. This fact means that we may overestimate
$\tau_{13}$ and $\tau_{12}$ by less than a factor of two, or we may
underestimate them by as much as a factor of five.

Since the $^{12}$CO lines are optically thick, the ALMA
$^{12}$CO(2--1) and $^{12}$CO(3--2) maps can be used to calculate the
radiation temperature of the gas by comparing the observed surface
brightness with a blackbody. For optically thick lines, the radiation
temperature equals the excitation temperature. Because the intrinsic
line width and the significant macroscopic movements of the gas as it
orbits the central star decrease the optical depth at a certain
frequency, we always selected the frequency channel with the maximal
intensity for each map pixel. This way, for each map pixel, we can
estimate the temperature at the frequency (or velocity) where the
optical depth is the highest. For $^{12}$CO we obtained excitation
temperatures up to 9\,K. The temperature can also be estimated from
the ratio of the optically thick $^{12}$CO(3--2) and $^{12}$CO(2--1)
lines. In the Rayleigh-Jeans approximation, the ratio is expected to
be the ratio of the squares of the line frequencies, i.e.~about
2.25. Instead, using the total flux values from Table
\ref{tab:results}, we obtain 1.16. The low value suggests that the
temperature is very low and that the Rayleigh--Jeans approximation is
not valid. Indeed, using the Planck function, it turns out that these
line ratios correspond to 5.6\,K.

In \citet{moor2011b}, given a lack of more information, we assumed
that $^{12}$CO was optically thin and estimated a total CO mass of
$3.5 \times 10^{-4}\,M_{\oplus}$. Our ALMA observations made it
evident that $^{12}$CO was optically thick and that C$^{18}$O should
be used for mass estimation. The calculations depend on the
temperature of the gas, for which we used different values between
5.6\,K (the excitation temperature obtained from the $^{12}$CO (3--2)
to (2--1) line ratio) and 100\,K (the hottest dust temperatures at the
inner edge of the disk in the model of Paper\,I). Higher temperatures
mean higher emission, but also mean that these low-$J$ transitions are
less excited. The net result of these two effects is that within the
assumed temperature range, the total CO mass needed to produce the
observed emission changes by only a factor of two. Thus, the CO mass
is well constrained, even without knowing the precise temperature.
The resulting value, (4--8)$\times$10$^{-2}\,M_{\oplus}$, is about two
orders of magnitude higher than previously believed. Considering the
uncertainties in the $^{12}$CO, $^{13}$CO and C$^{18}$O isotopic
ratios mentioned above, the total CO mass may actually be a factor of
two smaller or a factor of five higher than this value. Nevertheless,
the precise value does not change the fact that HD\,21997 has an
unusually gas-rich disk.

If the gas has a second-generation origin, the most common species,
based on cometary composition, are H$_2$O, CO, and CO$_2$ in
comparable abundances \citep[][and references
  therein]{mumma2011}. Thus, the CO mass gives a good estimate for the
order of magnitude of the total gas mass in the disk. However, if the
gas is primordial, then not only CO but also H$_2$ gas is present in
the disk and the total gas mass is significantly higher. Taking a
canonical CO/H$_2$ abundance ratio of 10$^{-4}$, the total gas mass is
on the order of 26--60$\,M_{\oplus}$. If such a large amount of H$_2$
is indeed present, and it is warm enough, the gas should display
rotational lines at mid-infrared wavelengths. We checked the {\it
  Spitzer}/Infrared Spectrograph spectrum of HD\,21997 (Paper\,I) for
the $S$(0), $S$(1), and $S$(2) lines, but found that they are not
present. Using a 3$\sigma$ upper limit for the flux of the line
expected to be the brightest, $S$(0), we calculated an upper limit of
35$\,M_{\oplus}$ for the total gas mass for 100\,K gas and
3400$\,M_{\oplus}$ for 50\,K gas. This result means that even if H$_2$
is present, it should be quite cold, just like the CO gas.

\subsection{Distribution of the CO Gas}
\label{sec:model}

\begin{table}
\begin{center}
{\small
\caption{Parameters of the HD\,21997 System and Its Gas Disk\label{tab:model}}
\begin{tabular}{ccc}
\tableline
Distance     & $d$ (pc)         & 71.95 \\
Systemic velocity & $v_{\rm LSR}$ (km\,s$^{-1}$) & 1.29 \\
Disk orientation & $P.A.$ ($^{\circ}$)   & $22.6 \pm 0.5$ \\
\tableline
Disk inner radius & $r_{\rm in}$ (AU)  & $< 26$ \\
Disk outer radius & $r_{\rm out}$ (AU) & $138\pm20$ \\
Stellar mass & $M_*$ ($M_{\odot}$) & $1.8^{+0.5}_{-0.2}$ \\
Disk inclination  & $i$ ($^{\circ}$)    & $32.6 \pm 3.1$ \\
Disk brightness exponent & $p$         & $-1.1 \pm 1.4$ \\
\tableline
\end{tabular}
}
\end{center}
\end{table}

\begin{figure}[h!]
\epsscale{1.18}
\plotone{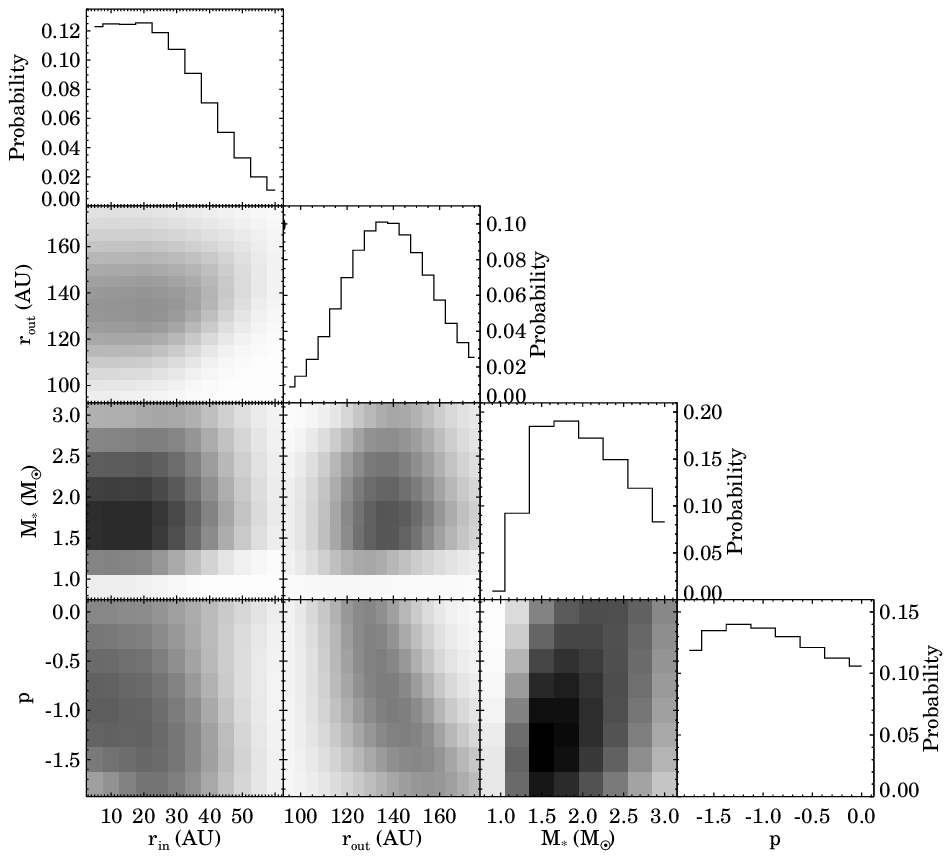}
\caption{Marginal probability distribution for the different model
  parameters. Details of the model are described in
  Section~\ref{sec:model}.
\label{fig:bayesian}}
\end{figure}

To determine the basic parameters of the HD\,21997 gas disk, we fit
the spatial and velocity distributions of the observed CO emission
with a simple disk geometry combined with a Keplerian velocity
profile. Our simple model has five parameters: the stellar mass
($M_*$), the inner and outer disk radii ($r_{\rm out}$, $r_{\rm in}$),
the inclination of the disk ($i$, where $i$=0$^{\circ}$ corresponds to
face on), and the exponent ($p$) of the radial brightness profile,
assumed to be a power law ($I(r)\,{\propto}\,r^{p}$). For the velocity
field of the gas, we adopted Keplerian rotation around the central
star, expected to be true for low-mass, non-self-gravitating disks
\citep[e.g.,][]{dutrey2007}. We used an intrinsic line width of
0.1\,km\,s$^{-1}$, corresponding to the Doppler broadening of a line
arising from $\approx$10\,K gas. We calculated the models for the same
velocity channels as the actual observations and applied a two-channel
wide Hanning smoothing to account for the instrumental velocity
resolution. In the spatial direction, we first calculated the models
with a 6$\times$ oversampling, convolved the models with the beam of
the actual observations, and rebinned them to the same pixel scale as
the observations. We decided to model the $^{12}$CO (2--1) line, which
is one of the brightest we observed and has the smallest and most
circular beam.

To explore a sufficiently large parameter range, we changed $M_*$
between 0.9$\,M_{\odot}$ and 3.0$\,M_{\odot}$ in steps of
0.3$\,M_{\odot}$, $r_{\rm in}$ between 5\,AU and 60\,AU in steps of
5\,AU, $r_{\rm out}$ between 95\,AU and 175\,AU in steps of 5\,AU, $i$
between 25$^{\circ}$ and 39$^{\circ}$ in steps of 2$^{\circ}$, and $p$
between 0 and $-1.75$ in steps of 0.25. In total, we calculated
104,448 models. In order to compare the models with the observations,
we normalized the models so that the total intensity equaled the
observed total intensity. Then, we computed the $\chi^2$ for each
channel map. The final $\chi^2$ of a certain model was defined as the
maximum of the $\chi^2$ of all channels. Finally, we calculated the
Bayesian probability for each model as $\exp(-\chi^2/2)$ and
normalized them so that the sum of the probabilities of all models is
1. Figure~\ref{fig:bayesian} shows the one-dimensional (1D)
probability distributions as a function of each parameter, and the
two-dimensional (2D) probability distributions as a function of two
different parameters. These are marginal distributions, i.e., for
example, in the distribution as a function of $r_{\rm in}$, all the
other parameters ($r_{\rm out}$, $M_*$, $i$, and $p$) were
marginalized out. Similarly, for example, in case of the 2D
distribution for $r_{\rm in}$ and $r_{\rm out}$, $M_*$, $i$, and $p$
were marginalized out (for more information on Bayesian inference and
marginalization, see, e.g., \citealt{loredo1992}). The value where the
1D probability distribution is maximal formally gives the best
estimate for that parameter. Formal 1$\sigma$ uncertainties can also
be estimated from these graphs by integrating the probability
distributions on each side of the maximum until one obtains
0.68. However, it is evident from Figure~\ref{fig:bayesian} that not
all parameters are equally well determined. The probability
distribution as a function of $r_{\rm in}$ is quite flat for short
inner radii, indicating that we can only give an upper limit on
$r_{\rm in}$. The outer radius $r_{\rm out}$ and stellar mass $M_*$
are well determined, while the power-law exponent of the brightness
profile, $p$, shows a very slowly changing and wide probability
distribution, which practically makes this parameter
ill-determined. The inclination and the stellar mass are degenerate,
thus we did not determine the inclination from the Bayesian
probabilities, but calculated it directly from the best-fit stellar
mass.

Table~\ref{tab:model} shows our best estimates for the disk parameters
and their 1$\sigma$ uncertainties, along with a 1$\sigma$ upper limit
for $r_{\rm in}$. Figure~\ref{fig:channelmaps} (right panel) shows
side-by-side the channel maps for the $^{12}$CO (2--1) observations,
the model with the best parameters (and $r_{\rm in}$=20\,AU), as well
as the residuals. Moreover, we plotted the integrated line profile
from the model in Figure~\ref{fig:cospectra}. These figures
demonstrate that the best model we found indeed fits the channel maps
and line profile very well. We convolved this model with a beam
appropriate for our other CO observations and scaled them so that the
integrated line intensity equals the observed intensities. We then
plotted the channel maps and residuals similarly to those in
Figure~\ref{fig:channelmaps} (right panel) and found that the
observations and the model match each other well. This fact
demonstrates that a single disk model with one set of parameters
reproduces all five CO lines.

In the following, we check whether the fitted parameters are
consistent with our a priori expectations. The major axis of the
$^{12}$CO(2--1) image is 4$\farcs$0 (Section~\ref{sec:resolved}),
which, deconvolved with the beam size of 1$\farcs$21
(Table~\ref{tab:results}), gives an outer radius of 137\,AU, in
agreement with our fitted value.  We can estimate the inclination from
the ratio of the minor and major axes of the $^{12}$CO(2--1) image,
which gives 29$^{\circ}$ (cf.~32$\fdg$6$\pm$3$\fdg$1 in
Table~\ref{tab:model}). A similar calculation was done from the minor
and major axes of the ellipse fit to the ALMA dust continuum image
(Paper\,I), giving 32$\fdg$9$\pm$2$\fdg$6. Another argument in favor
of a $\approx$30$^{\circ}$ inclination is that from the measured $v
\sin i$\,=\,70\,km\,s$^{-1}$ \citep{royer2007}, we obtain an
equatorial rotational velocity of 140\,km\,s$^{-1}$, typical for
A-type stars with a mass of about 1.8$\,M_{\odot}$
\citep{zorec2012}. Finally, the dynamical stellar mass of
1.8$\,M_{\odot}$ agrees well with the mass of 1.85$\,M_{\odot}$
obtained from evolutionary tracks by \citet{moor2011b}. These
comparisons demonstrate that our simple fitting procedure provided
reasonable results for the basic properties of the CO gas
distribution. A more detailed physical model, taking into account the
vertical disk structure and full radiative transfer, is postponed to
our forthcoming paper.


\section{Discussion}

Our modeling in Section~\ref{sec:model} suggests that the observed CO
emission can be well described by a gas disk in Keplerian rotation
around a 1.8$\,M_{\odot}$ central star. For the inner radius of the
gas disk, we obtained a 1$\sigma$ upper limit of 26\,AU. In Paper\,I,
we found that the dust disk starts at a radius of
$\approx$55\,AU. Consequently, we can claim with a 99\% confidence
that the inner radius of the gas is closer to the central star than
that of the dust disk. This result is in accordance with the
strikingly different appearance of the ring-like dust continuum image
and the centrally-peaked CO zeroth moment maps. There is a part in the
disk where the gas and the dust are not co-located and this ``inner
disk'' is practically dust-free. The total CO gas mass is about
(4--8)$\times$10$^{-2}\,M_{\oplus}$. In Paper\,I, we modeled the dust
emission of HD\,21997, and found that the dust mass within about
150\,AU is 0.09$\,M_{\oplus}$, resulting in a CO-to-dust ratio of
0.4--0.9, i.e., roughly the same mass of dust and CO gas is present in
the disk. This value may actually be slightly lower if we take into
account that a fraction of the gas is located in the inner dust-free
area. While the dust properties of the HD\,21997 system are
characteristic of a typical debris disk (Paper\,I), there is an
unexpectedly large amount of gas in the disk (for comparison, the
CO-to-dust ratio in $\beta$\,Pic is less than 2$\times$10$^{-4}$;
\citealt{dent1995,nilsson2009}). In the following, we discuss
different possibilities for the origin, physical properties, and
possible evolution of the gas component.

\subsection{Debris Disk with Pure Secondary Gas}
\label{sec:secondary}

In a debris disk, where neither dust grains nor H$_2$ provide enough
shielding for CO, the main factor determining the CO lifetime is its
self-shielding against the stellar UV radiation field and the
interstellar radiation field. To estimate the CO lifetime in the
HD\,21997 disk, we distributed the measured CO mass between 20\,AU and
138\,AU, adopted a power-law radial density distribution ($n \propto
r^{-\alpha}$, with $\alpha$\,=\,1.5, 2.0, and 2.5), and assumed three
different vertical scale heights ($H/r$\,=\,0.05, 0.1, and 0.2). We
computed the radial and vertical column density of CO in the disk and
estimated the shielding factors for $^{12}$CO and C$^{18}$O from the
tabulated values of \citet{visser2009}. We neglected the shielding
from the dust grains due to their low column density. We took into
account both stellar UV flux and the interstellar radiation field, the
same way as in \citet{moor2011b}. The resulting lifetimes are below
30,000\,yr for $^{12}$CO, and below 6000\,yr for C$^{18}$O. The
Keplerian rotation of the gas and resulting Doppler shift of the line
would further decrease the optical depth and reduce the
lifetime. Thus, a replenishment of CO molecules is needed.

Assuming an equilibrium between the dissociation and production of CO
molecules, and dividing the measured gas mass by the maximum lifetime
computed for $^{12}$CO, we obtained a lower limit for the gas
production rate of about 10$^{19}$\,kg\,yr$^{-1}$
(approx.~10$^{-6}\,M_{\oplus}$\,yr$^{-1}$). We emphasize that this
production rate is a very optimistic lower limit, because for a
significant part of the disk, CO lifetimes are lower than the value
quoted above. Typical gas production rates in solar system comets are
on the order of 10$^9$--10$^{10}$\,kg\,yr$^{-1}$
\citep{wyckoff1982}. Most of this gas, however, is H$_2$O, and CO
production is lower by a factor of 5--100 \citep{jewitt2007}. This
fact means that to keep up the observed CO mass in HD\,21997, the
continuous gas production of 10$^{10}$--10$^{11}$ comets is
needed. Alternatively, considering a larger comet like Hale--Bopp
(total mass of 1.3$\times$10$^{16}$\,kg; \citealt{weissman2007}) and
assuming a $\sim$10\% CO content \citep{sykes1986, mumma2011}, at
least 6000 Hale--Bopp-like comets need to be completely destroyed
every year. Unless one assumes a recent transient event in which a
large amount of CO was produced, the calculated very high
replenishment rate argues against the secondary origin of the CO gas
in the HD\,21997 system.

The destruction of planetesimals may result in both debris dust and
secondary gas (Section~\ref{sec:intro}). If the gas and dust indeed
have a common origin in the HD\,21997 system, one would expect that
they are co-located, unless some physical process separates gas and
dust and moves the gas inward or the dust outward from where they were
produced. One possibility is that the planetesimal ring starts about
55\,AU from the star, as the location of the dust disk
implies. Viscous accretion can transport material into the inner part,
but it would act both on gas and the small dust particles. The other
possibility would be to assume that the gas production via the
destruction of comets occurs in the warm inner disk. However, comets
would also produce dust, which was not observed. Thus, the existence
of the dust-free gas disk cannot be explained in the secondary
scenario.

\subsection{Primordial Gas in a Hybrid Disk}

The other possible scenario for the origin of the gas in the HD\,21997
system is that the CO molecules are remnants of the primordial
protoplanetary disk. Their survival would require very efficient
shielding over the 30 Myr long history of the system. Attributing this
shielding to the presence of H$_2$ gas (with a CO/H$_2$ abundance
ratio of 10$^{-4}$) and distributing the H$_2$ in the same simple disk
geometry as we assumed for the CO in Section~\ref{sec:secondary}, we
found CO lifetimes typically two orders of magnitude longer than
without the presence of H$_2$, long enough that some primordial gas
can still be present in the system. There is, however, one important
drawback of the primordial gas scenario. The calculation above shows
that the H$_2$ number density in the midplane is high enough ($n_{\rm
  H_2}\,{>}\,10^6$\,cm$^{-3}$) for the CO gas to be collisionally
excited and be in local thermodynamic equilibrium (LTE), using the
critical densities from \citet{kamp2001}. In this case, the excitation
temperature equals the kinetic temperature. Our CO observations
indicate very low excitation temperatures, much lower than the dust
temperature (64\,K, based on spectral energy distribution fitting done
by \citealt{moor2011b}). It is an open question what could cause such
a difference. The relatively warm dust temperature explains why the CO
does not freeze out onto the dust grains.

The oldest phases in primordial disk evolution are represented by
transitional disks. From a morphological point of view, HD\,21997
might give the impression of a transitional disk where gas exists
within the dust-free hole (for such examples, see \citealt{brown2009}
or \citealt{casassus2013}). However, we argue that HD\,21997 is not a
transitional disk because our results indicate that its dust content
is debris-like. This conclusion is supported by the following: (1) in
transitional disks, the gas-to-dust ratio is typically below the
interstellar value of 100 \citep{keane2013}, while in HD\,21997, using
the total gas mass including H$_2$, this ratio would be in the order
of 300--700. This result suggests that HD\,21997 is unusually
dust-poor, possibly because the majority of the dust is already locked
up in planetesimals. (2) The dust mass is only 0.1$\,M_{\oplus}$,
significantly lower than typical for transitional disks. (3) There are
no spectral signatures of small particles, either silicates or
polycyclic aromatic hydrocarbons. (4) The limited lifetime of dust
particles even in the presence of gas requires replenishment
(Paper\,I).

Based on these arguments, it is possible that we detected a hybrid
system where the gas is primordial while the dust is already
secondary. This result would also explain why the dust and gas are not
co-located. Because the dust component of HD\,21997 consists of large
grains (${>}\,6\,\mu$m; Paper\,I), the grains trace the location of
the planetesimal belt starting at about 55\,AU. The gas, however, is
the remnant of the primordial circumstellar material and thus may
fill the full disk. It is interesting to speculate about the further
evolution of the gas component in this system. It may be a stable
structure, or it may be in the phase of final disappearance. We note
that there are still unresolved issues with the hybrid disk
scenario. What happened to the primordial dust in the central
dust-free inner disk? Planetesimals already formed in the outer disk,
as evidenced by the debris dust, but why are they missing in the inner
disk where dynamical timescales are shorter? Is it possible that
solids rapidly accreted to planet masses without affecting the gas
component (by forming super-Earths rather than Jupiters)? The future
discovery of similar hybrid systems may help to answer these
questions. Some of the answers may also help to better understand the
evolution of some transitional disks, where the central clearing may
contain tenuous secondary dust, while the outer disks consist of
primordial gas and dust.


\section{Summary and Conclusions}

We presented ALMA CO line observations of the disk around the 30 Myr
old star HD\,21997. We detected the (2--1) and (3--2) lines of
$^{12}$CO and $^{13}$CO, and the (2--1) line of C$^{18}$O. The line
profiles and channel maps show that the gas is in Keplerian rotation
around the central star. We calculated a grid of simple gas disk
models with varying inner and outer radii, inclinations, stellar
masses, and radial brightness profiles of CO emission. Using Bayesian
probability analysis, we found that the best model gives $r_{\rm in} <
26$\,AU, $r_{\rm out}=138$\,AU, $M_*$=1.8$\,M_{\odot}$, and
$i=32\fdg6$, while the brightness profile is undetermined. The CO line
ratios and intensities suggest that the gas temperature is very low
(on the order of 6--9\,K). The total CO mass in the disk, as
calculated from the optically thin C$^{18}$O line, is about
(4--8)$\times$10$^{-2}\,M_{\oplus}$. Comparing our results with those
obtained for the dust component from the ALMA continuum observations
in Paper\,I, we concluded that, in terms of mass, a similar amount of
CO gas and dust is present in the disk. Interestingly, the gas and
dust in the HD\,21997 system are not co-located: there is an inner,
dust-free gas disk extending from closer than about 26\,AU to about
55\,AU from the star.

We discussed two possible scenarios for the origin of the gas in the
HD\,21997 disk. First we explored the possibility that the gas is of
secondary origin, mainly CO produced by planetesimals. In this case,
the sub-critical gas densities would lead to non-LTE conditions, which
might explain the low excitation temperatures. However, the short CO
lifetimes and the necessary high CO production rates exclude this
scenario. The other possibility is that the gas is primordial. In this
case, there is a large amount of H$_2$ gas is the disk, leading to
significantly longer CO lifetimes. A primordial origin would explain
the different locations of the gas and the dust. Based on our ALMA
observations, we propose that HD\,21997 is a hybrid system, where
primordial gas is accompanied by secondary debris dust. This fact
challenges the current paradigm of disk evolution, because the age of
HD\,21997 exceeds both the model predictions for disk clearing and the
ages of the oldest T\,Tauri-like or transitional gas disks in the
literature \citep{kastner2008}.

\acknowledgments

We thank the anonymous referee for useful comments that helped us to
improve the manuscript. This paper makes use of the following ALMA
data: ADS/JAO.ALMA\#2011.0.00780.S. ALMA is a partnership of ESO
(representing its member states), NSF (USA) and NINS (Japan), together
with NRC (Canada) and NSC and ASIAA (Taiwan), in cooperation with the
Republic of Chile. The Joint ALMA Observatory is operated by ESO,
AUI/NRAO and NAOJ. This work was partly supported by the grant
OTKA-101393 of the Hungarian Scientific Research Fund. This work is
based in part on observations made with {\it Herschel}, a European
Space Agency Cornerstone Mission with significant participation by
NASA. Support for this work was provided by NASA through an award
issued by JPL/Caltech.

{\it Facility:} \facility{ALMA}.

\bibliography{paper}{}

\end{document}